\documentclass{elsart3}
\usepackage{amssymb}
\usepackage{graphicx}
\journal{Physica C}
\begin{document}

\begin{frontmatter}

\title{Scaling of the microwave magneto-impedance in
Tl$_2$Ba$_2$CaCu$_2$O$_{8+\delta}$ thin films.}

\author[tre]{E. Silva\corauthref{cor}},
\corauth[cor]{Corresponding author} \ead{silva@fis.uniroma3.it}
\author[tre]{N. Pompeo},
\author[tre]{R. Marcon},
\author[Sap]{S. Sarti},
\author[Jena]{H. Schneidewind}
\address[tre]{Dipartimento di Fisica ``E.Amaldi'' and
Unit\`{a} CNISM, Universit\`{a} di Roma Tre, V. Vasca Navale 84, 00146
Roma, Italy}
\address[Sap]{Dipartimento di Fisica and Unit\`{a} CNISM,
Universit\`{a} ``La Sapienza'', 00185 Roma, Italy}
\address[Jena]{Institute for Physical High Technology Jena,
P.O.B.100239, D-07702 Jena, Germany}

\begin{abstract}
We present measurements of the magnetic field-induced microwave
complex resistivity changes at 47 GHz in
Tl$_2$Ba$_2$CaCu$_2$O$_{8+\delta}$ (TBCCO) thin films, in the ranges 58
K$<T<T_{c}$ and 0$<\mu_{0}H<$0.8 T. The large imaginary part
$\Delta\rho_{2}(H)$ points to strong elastic response, but the data
are not easily reconciled with a rigid vortex model.  We find that,
over a wide range of temperatures, all the pairs of curves
$\Delta\rho_{1}(H)$ and $\Delta\rho_{2}(H)$ can be collapsed on a pair
of scaling curves $\Delta\rho_{1}[H/H^{*}(T)]$,
$\Delta\rho_{2}[H/H^{*}(T)]$, with the same scaling field $H^{*}(T)$.
We argue that $H^{*}(T)$ is related to the loss of vortex rigidity due
to a vortex transformation.
\end{abstract}

\end{frontmatter}

% main text

Not too close to $T_{c}$ the vortex state microwave response is
dominated by vortex motion.  While the correct approach to the
frequency dependence of vortex motion is still a debated topic, it is
believed that at high enough frequencies each vortex probes only a
single potential well (due to, e.g., a defect) due to the very small
induced oscillations.  In this case the Coffey-Clem (CC) model is
effective in takeing into account vortex viscous motion, pinning
and creep \cite{cc}.  By lowering the frequency, vortices experience
large drags from their equilibrium positions, and they can interact
with each other and with several potential wells.  In this case the
nature and distribution of pinning centers becomes crucial, and
several vortex phases can arise \cite{ffh} with very different
transport properties.
\begin{figure}[htb]
\includegraphics [width=6.8cm]{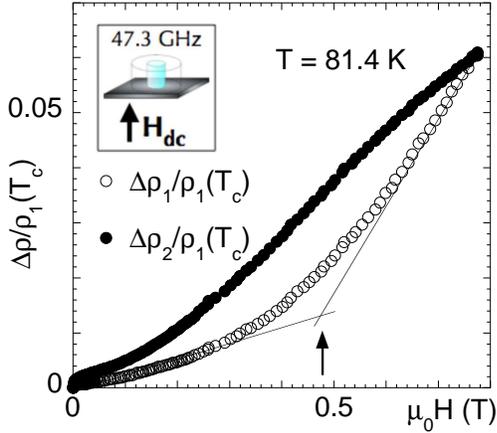} \caption{$\Delta\rho_{1}(H)$
(open symbols) and $\Delta\rho_{2}(H)$ (full symbols).  Arrow
indicates a possible choice for $H^{*}(T)$.  Inset: sketch of the
experimental configuration.  Microwave currents flow in the $ab$
planes.}
\label{fig1}
\end{figure}
In the single-vortex model the response is dictated by the upper
critical field, which determines the weight of the viscous flux flow,
and by (at least) a different characteristic field for vortex creep or
pinning.  By contrast, when the dynamics is dictated by a
transition or a crossover between different vortex phases instead of
pinning, there may be single characteristic field (in a sufficiently
small region of the $H,T$ phase diagram) describing vortex dynamics
\cite{ffh}.
In YBa$_2$Cu$_3$O$_{7-\delta}$ \cite{golos} a single-vortex model has
been applied with success to the description of the microwave
response.  However little is known about TBCCO.\\
We measured the microwave response at 47 GHz in TBCCO thin films by
means of a sapphire dielectric resonator operating in the TE$_{011}$
mode.
Changes in $Q$ factor and resonant frequency yielded changes in the
complex resistivity $\Delta\rho_{1}(H)+\mathrm{i}\Delta\rho_{2}(H)$.
We checked that, in the temperature range investigated, the thin film
approximation was valid.  The 240 nm-thick films have been grown on
2'' diameter CeO$_2$ buffered R-plane sapphire substrates by
conventional two-step method.  The resulting films show excellent
(100) orientation
and excellent in-plane epitaxy \cite{schneidewind}.  The
full-width-half-maximum of the $\theta-2\theta$ rocking curve is
0.4$^{\circ}$.  The film under study had $T_{c}\simeq$104 K and
$J_{c}=0.5$ MAcm$^{-2}$ measured inductively.\\
\begin{figure}[htb]
\includegraphics [width=6.2cm]{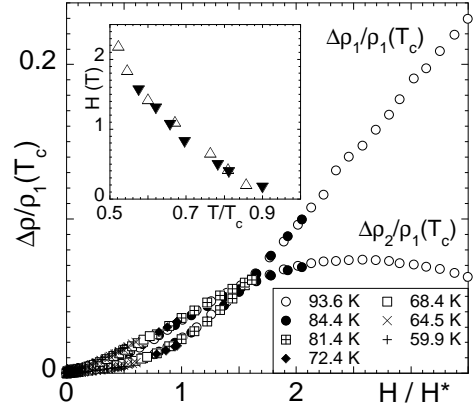} \caption{Collapse of
seven curves of $\Delta\rho_{1}$ and $\Delta\rho_{2}$ with $H^{*}(T)$.
For clarity, only 10\% of data is shown.  Inset: $H^{*}(T)$ (full
symbols) and melting field from \cite{ammor} (scaled by 2.2, open
symbols).}
\label{fig2}
\end{figure}
A typical measurement is reported in Fig. \ref{fig1}.  The large
imaginary part $\Delta\rho_{2}(H) \sim \Delta\rho_{1}(H)$ excludes
free-flux-flow, and indicates a strong elastic vortex response.
Moreover, the superlinear field dependence of $\Delta\rho_{1}(H)$ and
$\Delta\rho_{2}(H)$ is not compatible with a vortex motion dominated
by single-well strong pinning (Campbell regime).  However, a careful
analysis of the data within the periodic-potential CC model revealed
that they were incompatible also with significant creep.  In order to
reconcile the data with the CC model, a field-dependent depinning
frequency with no creep was needed, suggesting vortex collective
behaviour.  Due to the large amount of arbitrariness of this choice, we
concentrated in the identification of some more general, albeit less
model-specific, feature of the data. We found that, in agreement with the
indication of collective vortex behaviour,  our
measurements exhibited a clear field-dependent scaling, as
reported in Fig. \ref{fig2}: we found that
$\Delta\rho_{1}(H,T)+\mathrm{i}\Delta\rho_{2}(H,T)=
\Delta\rho_{1}[H/H^{*}(T)]+\mathrm{i}\Delta\rho_{2}[H/H^{*}(T)]$, with $H^{*}(T)$
that reproduced the temperature dependence of the vortex melting field in
TBCCO single crystals \cite{ammor} (inset of Fig.  \ref{fig2}),
suggesting that the dynamics was dictated by some vortex transformation.  While further
work is needed to clarify these issues, in this framework the large $\Delta\rho_{2}$
at low fields with the drop
at higher fields (see Fig.\ref{fig2}), and the temperature dependence 
of $H^{*}(T)$ would indicate a loss of vortex
rigidity at some vortex transformation instead of depinning of rigid
vortices.

\begin{thebibliography}{00}
%
\bibitem{cc} M. W. Coffey and J. R. Clem, {\it Phys. Rev. Lett.} {\bf 
67} (1991) 386
%
\bibitem{ffh} D. S. Fisher \etal, {\it Phys. Rev. B} {\bf 
43} (1991) 130; G. Blatter \etal, {\it Rev. Mod. Phys.} {\bf 
66} (1994) 115
%
\bibitem{golos} M.Golosovsky \etal, {\it Supercond. Sci. Technol.} {\bf 9} 
(1996) 1
%
\bibitem{schneidewind} H. Schneidewind \etal, {\it Institute of 
Physics Conf. Series} {\bf 167} (2000) 383
%
\bibitem{ammor} L. Ammor \etal, {\it Physica C} {\bf 282} (1997) 1983
%
\end{thebibliography}
\end{document}